\documentclass[letter]{aa}

\usepackage{graphicx}
\usepackage{txfonts}
\usepackage{lipsum}
\usepackage{subcaption}         
                            
\usepackage{lscape}             
                                
\usepackage{placeins}           
\usepackage{amsmath}	
\usepackage{relsize}
\usepackage{booktabs}
\usepackage{graphicx}
\usepackage{hyperref}
\usepackage{tablefootnote}
\usepackage{gensymb}
\usepackage{xcolor}

\hypersetup{
    colorlinks = true, 
    urlcolor = cyan, 
    linkcolor = blue, 
    citecolor = blue 
}
\usepackage{txfonts}
\usepackage{siunitx}
\usepackage{footnote}
\makesavenoteenv{figure}  
\usepackage{xcolor}
\usepackage{lscape}
\newcommand{\Msun}{\mbox{$\mathrm{M}_\odot$}}

\begin{document}

   \title{The youngest white dwarf companion to a millisecond pulsar: Insights from NGC~362D}

   \author{Greta Ettorre
          \inst{1,2},
          Emanuele Dalessandro
          \inst{2},
          Mario Cadelano
          \inst{1,2},
          Alessandro Ridolfi
          \inst{3}, 
          Cristina Pallanca
          \inst{1,2},\\
          Paulo C. C. Freire
          \inst{4}, 
          Vivek Venkatraman Krishnan
          \inst{4}, 
          Maurizio Salaris
          \inst{2},
          Franca D'Antona
          \inst{5},
          Rouhin Nag
          \inst{6}
          }

   \authorrunning{Ettorre et al.}         
   \institute{Department of Physics and Astronomy “Augusto Righi”, University of Bologna, Via Gobetti 93/2, 40129 Bologna, Italy
         \and
             INAF – Osservatorio di Astrofisica e Scienza dello Spazio, Via Gobetti 93/3, 40129 Bologna, (BO), Italy
        \and
            Fakultät für Physik, Universität Bielefeld, Postfach 100131, D-33501 Bielefeld, Germany
        \and 
            Max-Planck-Institut f\"ur Radioastronomie, auf dem H\"ugel 69, 53121, Bonn, Germany
        \and
            INAF - Osservatorio Astronomico di Roma, Via Frascati 33, 00078 Monte Porzio Catone, (RM), Italy
        \and
            INAF – Osservatorio Astronomico di Cagliari, Via della Scienza 5, I-09047 Selargius, (CA), Italy
}
   \date{Received XXX; accepted YYY} 
 
  \abstract
   {We report on the identification of the optical counterpart to the recently discovered millisecond pulsar (MSP) NGC362D in the Galactic globular cluster NGC~362
   based on deep, multi-band, and multi-epoch \textit{Hubble} Space Telescope observations. 
   Our analysis robustly shows that this object is a very low-mass ($\sim0.18\,\Msun$) He white dwarf (WD) still in the pre-cooling phase. Interestingly, a detailed comparison with updated binary evolution models indicates that this object completed the mass-transfer phase only recently ($\sim0.6$ Gyr ago), thus strongly suggesting that COM-NGC362D is the youngest WD companion to a MSP identified to date. 
   Remarkably, in this respect, the photometric properties of NGC362D show, for the first time in this class of objects, significant wavelength-dependent variations that are consistent with the presence of residual circumstellar material.  
   This system therefore provides a valuable test case to directly probe the immediate aftermath of the MSP recycling process and to constrain the early evolutionary stages of proto-WD companions and their radio and optical properties. Our results suggest that residual material can produce radio and optical signatures that mimic those of systems with non-degenerate companions, potentially leading to the misclassification of young MSPs.}

   \keywords{globular clusters: individual (NGC~362) --
                pulsars: general -- stars: evolution -- binaries: general
               }

   \maketitle

\nolinenumbers
\section{Introduction}
Millisecond pulsars (MSPs) form in binary systems containing a slowly 
rotating neutron star (NS) that is eventually spun up to millisecond periods 
by heavy-mass and angular-momentum transfer from an evolving 
companion \citep{Alpar1982,Bhattacharya1991,Tauris1999,DantonaTailo2022,Tauris2011}. 

The recycling scenario for MSP formation predicts (e.g. \citealt{Tauris2011}) that the mass transfer between the non-degenerate companion and the NS typically 
starts when the donor is either a main sequence (MS -- case A) or a subgiant or red-giant branch star (case B).
During the early recycling phases, the binary system is observable as a low-mass X-ray binary (LMXB), and
then it progressively evolves into a purely rotation-powered radio MSP \citep{Wijnands1998,Archibald2009,Papitto2013,DiSalvo2020,Patruno2021}
and the non-degenerate companion is expected to progressively approach the white dwarf (WD) cooling sequence.
Alternatively, some MSPs are found in binaries with low-mass, non-degenerate companions. Depending on the companion mass, these systems are classified as redbacks (RBs) or black widows (BWs). The formation and evolutionary scenario of these systems is still strongly debated (e.g. \citealt{chen2013,Benvenuto2014,Benvenuto2015,Smedley2015,Devito2020}) 
and appears to be linked to the action of the strong MSP wind on the companion star.
In all of these cases, however, the evolutionary state of the companion star and the orbital period of the system at the onset of 
the Roche lobe overflow (RLO) are the determining factors for the outcome of the mass transfer and the properties of the pulsar itself \citep{Tauris2011}.

\begin{figure*}
    \centering
    \includegraphics[width=0.9\textwidth]{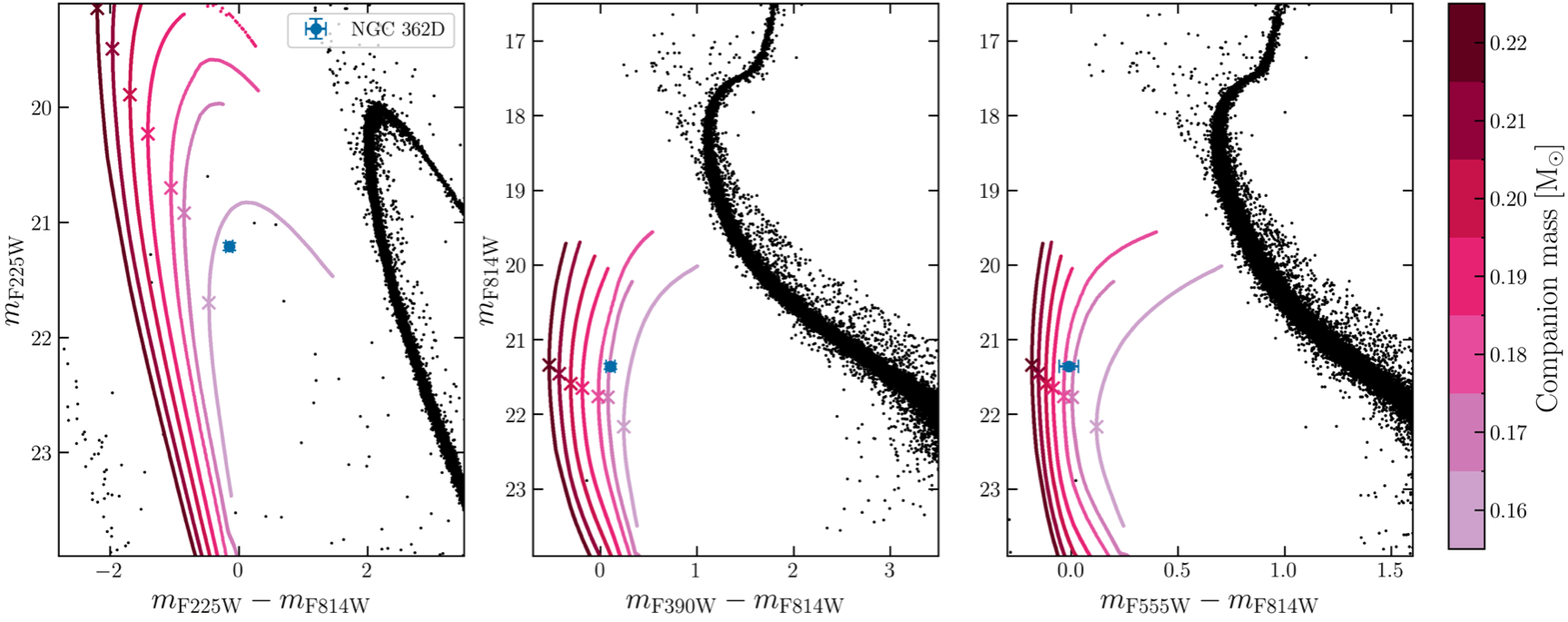}
    \caption{\textit{Left}: Observed ($m_{\rm F225W}$, $m_{\rm F225W} - m_{\rm F814W}$) colour–magnitude diagram. Black points represent cluster-member stars from our catalogue with good photometric quality. The optical counterpart to NGC362D is marked with a blue point, together with the error bars reflecting the corresponding photometric uncertainty. Overplotted are the He WD cooling tracks used in this work. These correspond to stellar masses between $0.16\,\Msun$ and $0.22\,\Msun$, as indicated by the colour bar. For each track, the beginning of the cooling phase is marked with a cross. \textit{Middle and right}: Same CMDs as shown on the left, but in the ($m_{\rm F814W}$, $m_{\rm F390W} - m_{\rm F814W}$) and ($m_{\rm F814W}$, $m_{\rm F555W} - m_{\rm F814W}$) filters, respectively.}
    \label{fig2}
\end{figure*}
While the recycling scenario succeeds in reproducing many properties of the observed population of MSPs and their companions,   
many details of the mass-transfer and accretion processes, as well as their associated timescales, remain uncertain. In particular, while the initial stages of mass accretion, the resulting end products and the connection between LMXBs and MSPs are relatively well understood both in terms of theoretical predictions and observations, we still lack an adequate sampling of the intermediate evolutionary phases, when these systems are still transitioning from an active mass-transfer phase to a more stable MSP evolution. For example, the residual hydrogen burning at the bottom of the companion's envelope can keep the WD warm for $\sim10^9$ years, leading to extremely long timescales of proto-WD evolution \citep{Alberts1996,Althaus2001b,vanKerkwijk2005}, with non-negligible implications on how MSP ages compare when inferred by means of characteristic spin-down timescales from radio observations \citep[see][]{Tauris2012} or WD cooling ages. Also, the poorly constrained properties of MSPs during these transitional phases, both at radio and optical wavelengths, may lead to puzzling observational behaviours (see, e.g. the well-known case of PSR J1816+4510; \citealt{Kaplan2012,Kaplan2013,Polzin2020,Shang2024}) and can ultimately result in an uncertain classification of such systems.  
In this context, the identification and full characterisation of optical counterparts to MSPs in these intermediate evolutionary phases, is essential to fully constrain their evolutionary pathways and eventually to obtain a better understanding of their radio and X-ray properties \citep[e.g.][]{Edmonds2002,Ferraro2003}.

As part of a long-standing multi-wavelength campaign to characterise the optical companions of MSPs (see \citealp{Ettorre2025a} and references therein),
in this letter we report on the identification and characterisation in terms of mass, 
effective temperature, cooling age, and mass-transfer scenario 
of the companion star to the MSP NGC362D (PSR J0103$-$7050D) in the Galactic globular cluster (GC) NGC~362. 
This is a binary MSP, which was recently identified by using MeerKAT observations (Ridolfi et al. in prep.) as part of the TRAPUM survey. Interestingly, we find that the optical counterpart to NGC362D is the youngest WD companion to a MSP observed so far, thus potentially representing a useful benchmark to probe the properties of MSPs during their intermediate evolutionary stages.

\section{Identification and photometric characterisation of COM-NGC362D}
In this letter, we make use of the extensive multi-wavelength and multi-epoch \textit{Hubble} Space Telescope (HST) dataset presented in \citet{Ettorre2025b}, where all details regarding the photometric reduction, the construction of the final catalogue, and the membership selection are fully described. Here, we briefly note that the dataset consists of images obtained in nine different filters, spanning from the near-UV (F225W) to the optical I band (F814W). The observations were obtained over a time baseline of approximately ten years, from 2006 to 2016. 
The timing solution for NGC362D was obtained by Ridolfi et al. (in prep.).
While we refer the reader to that paper for a full description of the radio observations, data, and timing analysis, we note here that NGC362D is found to be a binary system in a circular orbit of 0.17 days, with a companion minimum mass of $0.11\,\Msun$ (assuming a pulsar mass of $1.4\,\Msun$). 
These properties are consistent with evolution through the case A evolutionary channel and, together with the absence of radio eclipses, support the interpretation that the companion is a low-mass He WD \citep{Tauris2012}.

We searched for the optical companion to NGC362D by inspecting all the available images and relative catalogues within a radius of $0.1''$ (see Appendix~\ref{appendix1} for details) from the pulsar position (Ridolfi et al. in prep.), after shifting it to the epoch of the optical observations by accounting for the proper motion and assuming that the pulsar moves with the bulk of the cluster stars. 
We identified a single source within the search radius, located at only $0.02''$ from the pulsar (see the finding chart in Appendix~\ref{appendix1}). As we detail below, the properties of this star nicely match the 
expectations from radio studies, thus making it a good candidate to be the optical counterpart to NGC362D. Hence, hereafter we 
label this star as COM-NGC362D. 

The star COM-NGC362D is a relatively bright cluster member ($m_{\rm F555W}=21.3$ mag; see Appendix~\ref{appendix2}) detected in all the available filters in our catalogue. In all colour--magnitude diagrams (CMDs) COM-NGC362D lies in the region between the CO-WD cooling sequence and the cluster MS (Fig.~\ref{fig2}), and its position is compatible with He-WD cooling tracks (see Sect.~\ref{section4}).

\begin{figure}
    \centering
    \includegraphics[width=0.95\hsize]{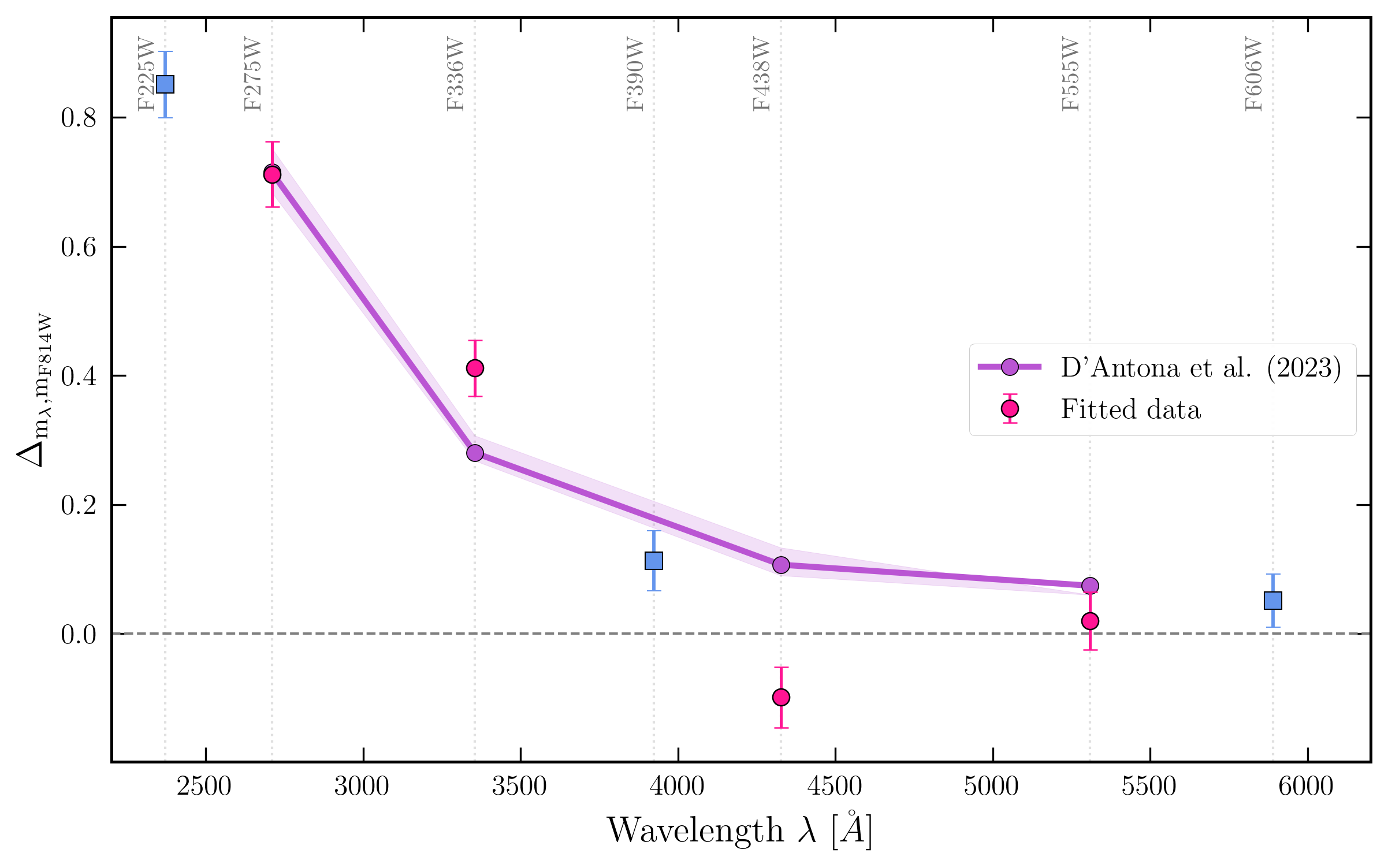}
    \caption{Colour difference, $\Delta_{\mathrm{m}_\lambda,\mathrm{m_{F814W}}}$, between the observed $(\mathrm{m}_\lambda-\mathrm{m_{F814W}})$ colour of the NGC362D companion and the colour predicted by the reference $0.18\,\Msun$ He WD model, plotted as a function of wavelength. Pink points represent the fitted data, while blue squares mark measurements not used in the fit. The purple line shows the prediction from Eq.~(1) of \citet{DAntona2023}, with the shaded region indicating the associated uncertainty. The observed trend, with larger deviations at shorter wavelengths, is consistent with a self-extinction effect likely produced by circumstellar material surrounding the proto-WD companion.}
    \label{fig4}
\end{figure}

Consistently with COM-NGC362D being a WD, we find no evidence of photometric variability for this object in any of the available filters (see Appendix~\ref{appendix3}). It is worth stressing here that photometric variability is expected for non-degenerate companions 
(as in RB systems), while it is uncommon for degenerate companions and, in the cases where it is observed, it is usually not driven 
by irradiation from the MSP, but rather by intrinsic processes within the WD, such as global stellar pulsations \citep{Maxted2013,Kilic2015}.
Finally, following the same procedure described in \citet{Ettorre2025b} and widely used in previous works \citep{Beccari2013,Pallanca2013,Pallanca2017}, 
we find that COM-NGC362D does not show any significant H$\alpha$ excess \citep[see Figure 8 in][]{Ettorre2025b}.
This is expected as NGC362D is detected as a radio pulsar and does not show any strong X-ray emission \citep{Kumawat2024,Ettorre2025b}, consistent with a system that has likely already completed the mass-transfer phase.

\section{Physical properties of COM-NGC362D}\label{section4}
A close inspection of Fig.~\ref{fig2} shows that the position of COM-NGC362D is consistent with the pre-cooling portion\footnote{This is the evolutionary phase starting at the end of the RLO and ending at the beginning of the cooling, when the $T_{\rm eff}$ of the WD reaches its maximum. The onset of the cooling phase is marked as a cross on each track in Fig.~\ref{fig2}.} of the He-WD tracks by \citeauthor{Istrate2016} (\citeyear{Istrate2016}; see also \citealp{Cadelano2019,Cadelano2020}, for a full description of the models),
thus suggesting that COM-NGC362D is still a proto-WD and the mass-transfer process with the NS likely ended only recently. 
As a reference, we used the magnitudes $m_{\rm F555W}$ and $m_{\rm F814W}$ to fit the observed CMD position of COM-NGC362D using the aforementioned He-WD cooling models. To place the He-WD cooling tracks onto the observed CMD, we adopted a colour excess of $E(B-V)=0.05$ \citep{Harris2010} and a distance modulus of $\mu=14.73$ \citep{Baumgardt2021}. The extinction in each filter was computed using the extinction law of \citet{Cardelli1989}, assuming $R_V = 3.1$.
To perform the fit, we adopted the same approach described in \citet{Ettorre2025a}, which uses a multivariate Gaussian logarithmic likelihood. 
The likelihood distributions of the companion's parameters are shown in Fig.~\ref{fig5app} (see also Appendix~\ref{appendix5}).
We find that the position of COM-NGC362D in the ($m_{\rm F814W}$, $m_{\rm F555W}$-$m_{\rm F814W}$) CMD is consistent with that of a He-WD with a mass of $0.18^{+0.01}_{-0.01}\,\Msun$ and a surface temperature of $\log\mathrm{T}_\mathrm{eff} = 4.11^{+0.05}_{-0.03}$ K---still in the proto-WD phase---$180^{+90}_{-80}\,\mathrm{Myr}$ before the onset of the cooling stage. 
According to Eq.~1 of \citet{Istrate2014}, for a WD companion of this mass the proto-WD phase lasts $\sim0.87$ Gyr, implying that the Roche-lobe 
detachment occurred only $\sim0.69$ Gyr ago and thus strongly suggesting that COM-NGC362D is the youngest WD companion to a MSP identified to date.

In this respect, the derived cooling age does not change significantly when using different sets of low-mass He-WD models \citep{Althaus2013}. Moreover, the null eccentricity inferred via radio timing supports the hypothesis that the recycling process concluded only recently, leaving insufficient time for dynamical interactions with neighbouring stars in the GC to perturb the orbit.
We note, in passing, that NGC362D would therefore increase the known population of MSP–He WD binaries with $\mathrm{P_{orb}}<1$ day. The existence of a growing number of such systems reinforces the tension noted by \citet{Istrate2014b}, whose LMXB models require a significant fine-tuning of the initial orbital period to reproduce detached He WDs in these tight orbits.

Interestingly, we observe that the position of COM-NGC362D with respect to the track of the best-fit mass in the ($m_{\rm F555W} - m_{\rm F814W}$) CMD changes as a function of the adopted colours in the three panels of Fig.~\ref{fig2}, becoming progressively redder when moving to shorter wavelengths. 
We can safely rule out that the colour drift observed for COM-NGC362D can be caused by time variability, such as instability strip–like pulsations \citep{Kilic2015} or the source being caught during a hydrogen-shell flash \citep{Althaus2001b,Istrate2014,Istrate2016}, which could produce inconsistent positions in CMDs built with filters from different epochs (i.e. at random phases; see Appendix~\ref{appendix4}). 

Our proposed scenario envisages that COM-NGC362D is an extremely young WD that only recently ended the mass-transfer phase and is surrounded by a residual cloud of material producing self-extinction. This is qualitatively similar to what was proposed by \citet{DAntona2023} 
to reproduce the observed properties of UV-dim stars in young massive clusters in the Magellanic Clouds (see also \citealt{Kamann2023,Leanza2025}). One possibility is that the circumstellar material is a remnant of the mass-transfer process, but admittedly it is unclear whether it can survive around the companion for $\sim0.6$ Gyr after the RLO detachment. Alternatively, the material may have been expelled during a previous hydrogen-shell flash. In this scenario, each flash is expected to eject $\sim10^{-4}\,\Msun$ of material over a timescale of $\sim10^3$ yr \citep{Antoniadis2014}. However, the expelled material would also be expected to induce a measurable orbital eccentricity on short timescales ($10^4$--$10^5$ yr), which is at odds with the currently observed circular orbit of NGC362D.

To quantitatively test the self-extinction hypothesis, we computed the $\Delta_{\mathrm{m}_\lambda,\mathrm{m_{F814W}}}$ residual, which is defined as the difference between the 
observed $(\mathrm{m}_\lambda-\mathrm{m_{F814W}})$ colour and that predicted by the $0.18\,\Msun$ reference model, as a function of wavelength, 
following the same approach adopted by \citet{Leanza2025}. Figure~\ref{fig4} shows that $(\mathrm{m}_\lambda-\mathrm{m_{F814W}})$ varies monotonically and by up to $\sim0.8$ mag when moving from the bluest to the reddest available filter. 
Such a wavelength-dependent trend  
is at least qualitatively consistent with the presence of self-extinction affecting the companion star \citep{DAntona2023}. 
As a sanity check, we verified that the application of the same analysis to the available sample of CO WDs yields constant $\Delta_{\mathrm{m}_\lambda,\mathrm{m_{F814W}}}$ (within the uncertainties) for all
filters (see Fig.~\ref{fig6app}).
For a more quantitative analysis, we applied Eq.~1 from \citet{DAntona2023}, which expresses the dust-absorbed magnitude of a star affected by self-extinction due to circumstellar material as a function of the optical depth, $\tau_{10}$, at $10\,\mu$m and the inclination angle, $i,$ of the disc to the line of sight. 
The fit was performed using only the HST bands for which the absorption coefficient $\delta m(\tau_{10})$ can be extrapolated from \citet{DAntona2023}; these are shown as pink points in Fig.~\ref{fig4}. The remaining bands (blue squares) were excluded from the fit, as the corresponding absorption coefficients are not available.\footnote{With the starting grid values from \citet{DAntona2023}, the best-fit grain size and optical depth turn out to be $0.05\,\mu\mathrm{m}$ and $\tau_{10} = 0.18 \pm 0.01$, respectively, when assuming an inclination $i=60^{\circ}$.}
Most of the data points are well reproduced by the model’s predicted trend (shown in purple), with the exception of the $m_{\rm F438W}$ band, which shows a more pronounced deviation compared to the other bands. Overall, the agreement between the observations and the model supports the idea that COM-NGC362D is still surrounded by some residual material.

\section{Discussion and conclusions}
\label{sec:discussion}
In this letter we report on the identification and characterisation of the optical counterpart to the binary MSP NGC362D using deep, multi-band, and multi-epoch HST observations. 
Interestingly, COM-NGC362D turns out to be the youngest WD companion to a MSP known to date.
Moreover, its photometric properties and its wavelength-dependent variations strongly suggest the presence of residual circumstellar material likely expelled during the mass-transfer phase. 
Regardless of the exact origin of the material, which remains an open question, this system provides a test case to directly probe the immediate aftermath of the MSP recycling process and to constrain the early evolutionary stages of proto-WD companions and their radio and optical properties. 

In this respect, we note that some of the properties inferred for COM-NGC362D resemble those observed in the enigmatic Galactic-field system PSR~J1816+4510 \citep{Kaplan2012,Kaplan2013,Polzin2020,Shang2024}. This system was initially classified as a RB \citep{Kaplan2012,Stovall2014} because of the presence of radio eclipses lasting for $7\%$--$10\%$ of the orbital period. However, optical spectroscopic studies \citep{Kaplan2013} revealed a companion with a minimum mass of $0.19\,\Msun$, $\mathrm{T_{eff}} = 16000$\,K and low surface gravity, which are consistent with 
expectations for a proto-WD. \citet{Polzin2020} reported a transition in the eclipsing mechanism of PSR~J1816+4510, from direct flux removal to a simple smearing of the radio signal, which was interpreted as evidence of the presence of an extended tail of material. Moreover, \citet{Kaplan2013} detected narrow He\,\textsc{i}, Ca\,\textsc{ii}, Si\,\textsc{ii}, and Mg\,\textsc{ii} lines, which could originate in a lower pressure environment possibly associated with this cloud of expelled material. 
Therefore, it is very tempting to conclude that PSR J1816+4510 and COM-NGC362D share 
very similar properties and, very likely, a similar evolutionary stage.

We can thus argue that residual circumstellar material may persist around the companion star for a significant fraction of the pre-cooling timescale, potentially producing observable signatures at radio and optical wavelengths that are not predicted by standard evolutionary models and that can ultimately lead to the ambiguous or incorrect classification of such systems.

Within this interpretative framework, the absence of radio eclipses in NGC362D (Ridolfi et al. in prep.) may be explained by the inhomogeneous and clumpy distribution of the surrounding material, as observed by \citet{Polzin2020} and \citet{Shang2024} for PSR~J1816+4510 from the detection of dispersion measure fluctuations on short spatial and temporal scales. Alternatively, the expelled material may have a column density below the threshold required to produce detectable radio absorption or scattering at the observed frequencies.
Finally, we also note that the optical companion to the MSP M2C \citep{Li2025} appears to be very similar to COM-NGC362D. In fact, it is a young He-WD, at about 10 Myr since the beginning of its cooling, possibly showing  a similar wavelength-dependent colour variation; this may suggest that it has been caught in the same transitional post-mass-transfer phase as COM-NGC362D.

Given the brightness of COM-NGC362D, future spectroscopic observations could be used to measure the radial-velocity curve of the companion star, thereby placing strong constraints on the binary mass ratio. Moreover, the detection of narrow metal lines with profiles similar to those observed in PSR~J1816+4510 \citep{Kaplan2013} or by \citet{Leanza2025} for the case of UV-dim stars, would provide further support for the interpretative scenario we describe in this letter.

\begin{acknowledgements}
G.E. acknowledges support from the Next Generation EU funds within the National Recovery and Resilience Plan (PNRR), Mission 4 - Education and Research, Component 2 - From Research to Business (M4C2), Investment Line 3.1 - Strengthening and creation of Research Infrastructures, Project IR0000034 – ``STILES - Strengthening the Italian Leadership in ELT and SKA''. GE and ED acknowledge financial support from the INAF Data analysis Research Grant (PI: E. Dalessandro) of the “Bando Astrofisica Fondamentale 2024”.
\end{acknowledgements}

\bibliographystyle{aa} 
\bibliography{biblio} 

\begin{appendix}
\section{The search for COM-NGC362D}\label{appendix1}
\begin{figure*}
    \centering
    \includegraphics[width=\hsize]{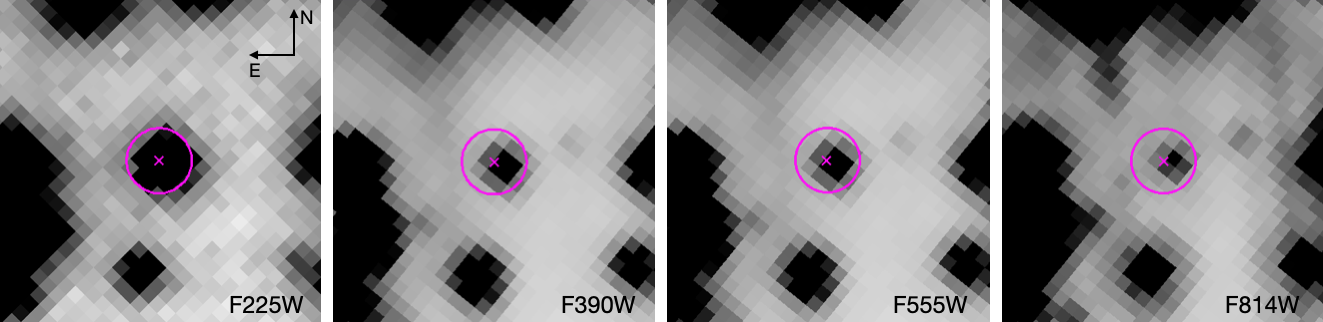}
    \caption{Finding charts for COM-NGC362D. The panels display \texttt{\_drc} images in the following HST filters, from left to right: WFC3/UVIS F225W (GO-14155), F390W, F555W, and F814W (GO-12516).  Each panel shows a $1\arcsec \times 1\arcsec$ FoV centred on the radio position of the pulsar. The magenta cross marks the nominal position of the MSP, while the magenta circle (radius $0.\!\arcsec1$) indicates the search radius. North is up, and east is to the left.}
    \label{fig2app}
\end{figure*}

In Fig.~\ref{fig2app} we present the finding chart of COM-NGC362D in the HST/WFC3 F225W (proposal GO-14155), F390W, F555W, and F814W (proposal GO-12516) filters. The magenta cross indicates the radio position of the MSP NGC362D, while the magenta circle, with a radius of $0.\!\arcsec1$, defines the search region used in this work to identify the optical counterpart. The size of the search radius is driven by the astrometric uncertainties of the optical catalogue, which are due to the residuals of the polynomial transformations used to report the HST instrumental coordinates to the Gaia reference frame. 

\section{Membership of COM-NGC362D}\label{appendix2}
\begin{figure}
    \centering
    \includegraphics[width=\hsize]{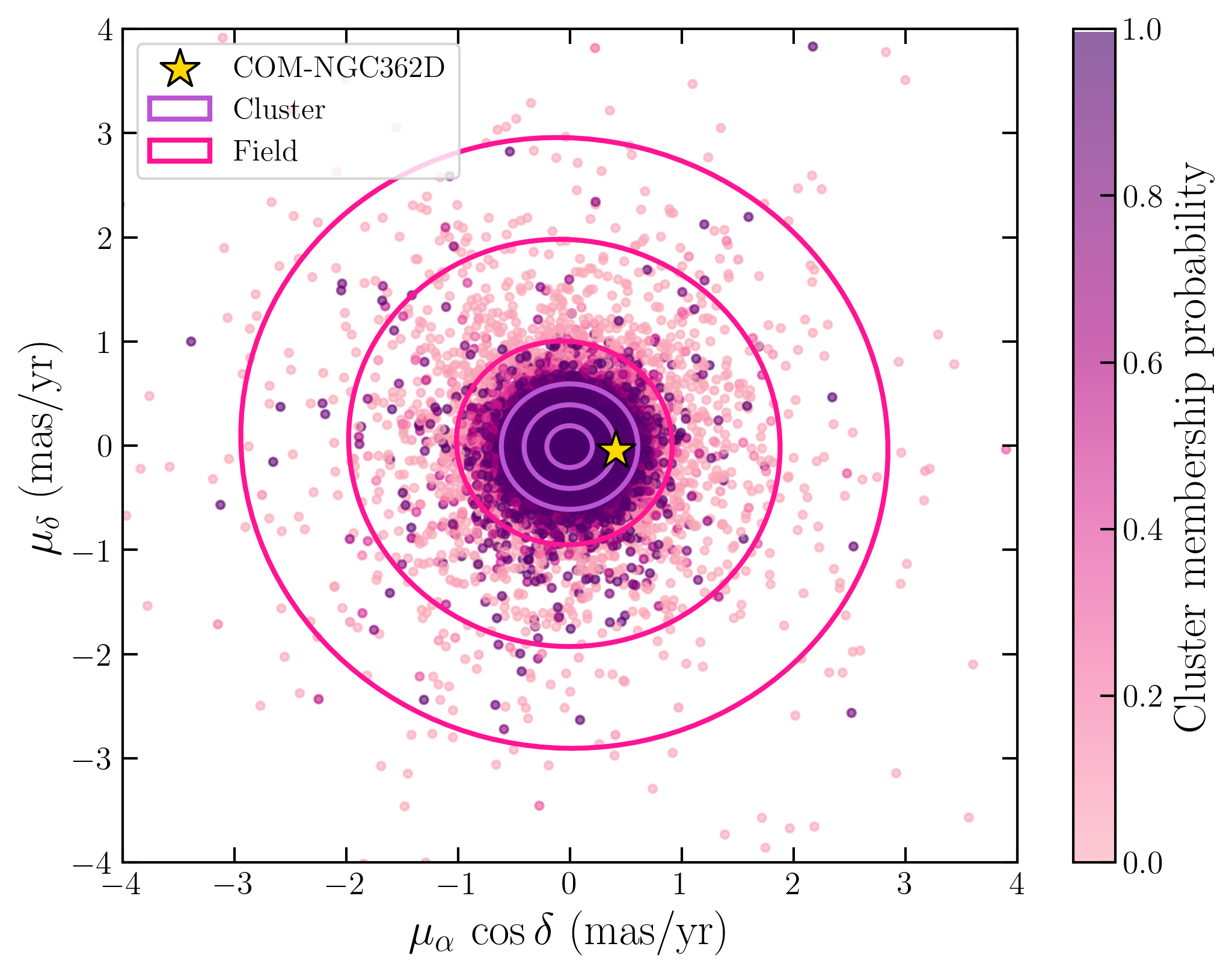}
    \caption{Proper motion VPD for stars within the F606W magnitude bin that includes COM-NGC362D (shown with a yellow star). The two Gaussian components, representing the cluster and field populations respectively, are shown as purple and magenta ellipses (at 1, 2, and 3$\sigma$).}
    \label{fig3app}
\end{figure}

COM-NGC362D broadly overlaps in the CMDs with the distribution of the 
young MS of the background Small Magellanic Cloud (SMC) galaxy.
Therefore we investigated its cluster membership by using proper motions from the HSTPROMO catalogue. We selected stars within $\pm1$ mag in F606W of the magnitude of COM-NGC362D and we show their distribution in the Vector Point Diagram (VPD) in Fig.~\ref{fig3app}. The SMC population, centred at ($\mu_{\alpha}\cos\delta = -6$ mas/yr, $\mu_{\delta} = 1.5$ mas/yr) lies well outside the plotted region 
\citep[see the left panel of Fig. 3 in][]{Ettorre2025b}. 
The position of COM-NGC362D (gold star) appears to be compatible with that of cluster members (Fig.~\ref{fig3app}) and we can already exclude that it can be linked to the SMC. 
For a quantitative assessment of its membership, we modelled the observed distribution of stars in the 
VPD using a two-component Gaussian Mixture Model implemented through the XDGMM Python framework \citep{holoien2017}.
The two Gaussian components, representing the cluster and the Galactic field populations respectively, are shown as purple and magenta ellipses (at 1, 2, and 3$\sigma$). Individual stars are colour-coded by their resulting cluster membership probability. COM-NGC362D falls within the $\sim2\sigma$ contour of the cluster component, and is assigned a membership probability of $\rm P_{\rm memb} =0.99$, confirming its membership to NGC~362. 

\section{Photometric variability of COM-NGC362D}\label{appendix3}
\begin{figure}
    \centering
    \includegraphics[width=\hsize]{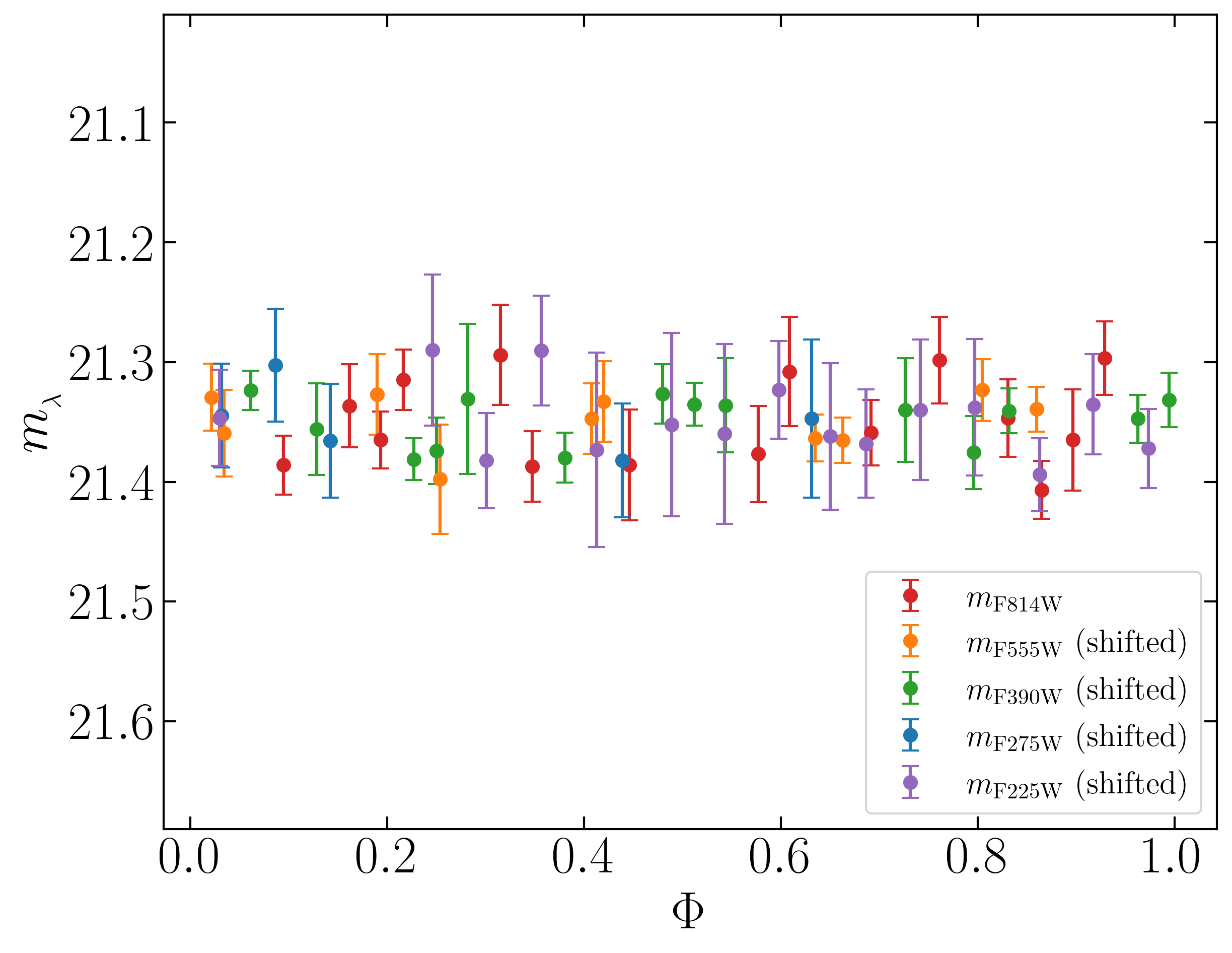}
    \caption{Global light curve of COM-NGC362D obtained by combining measurements from the F225W, F275W, F390W, F555W, and F814W WFC3 images (shown in purple, blue, green, orange, and red, respectively). The light curve is phase-folded using the 0.17-day period derived from radio timing. The F814W band was adopted as the reference magnitude, and the measurements in the other filters were shifted accordingly to construct the global light curve.}
    \label{fig4app}
\end{figure}

To study the variability of COM-NGC362D, we followed the same procedure described in detail in \citet{Ettorre2025b}. 
We used all the filters processed in our \texttt{DAOPHOT} photometric reduction to construct the global light curve, by shifting each filter by the offset between its mean magnitude and the mean magnitude in the F814W band, which was chosen as the reference. The global light curve is shown in Fig.~\ref{fig4app}.  The object shows a scatter around $m_{\rm F814W}\sim21.35$ that is fully consistent with the photometric uncertainties, indicating no detectable intrinsic variability. 

\section{Derivation of the main properties of COM-NGC362D}\label{appendix5}
\begin{figure*}
\sidecaption
\includegraphics[width=12cm]{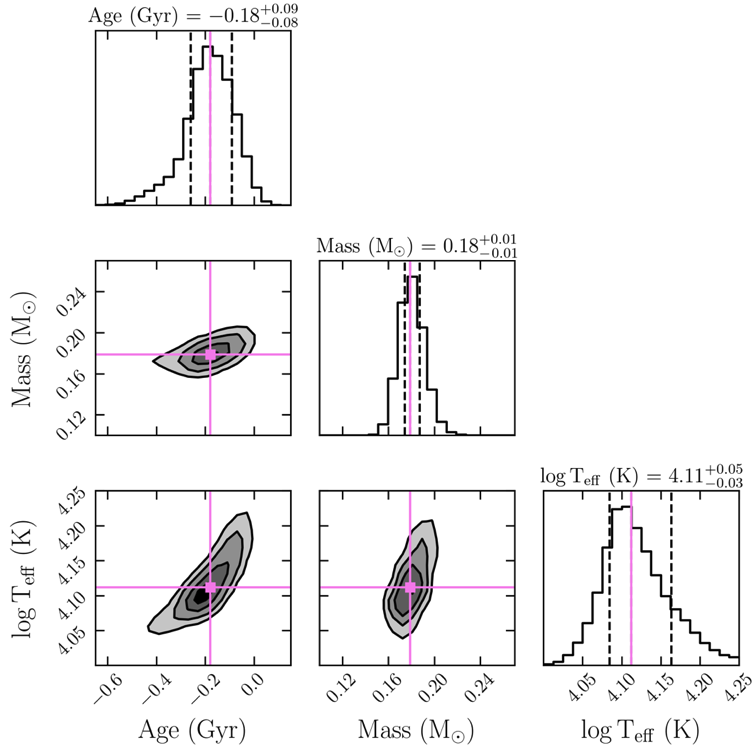}
\caption{Constraints on the mass, cooling age, and effective temperature of the companion star to NGC362D. The 1D histograms show the marginalized likelihood distributions for each parameter, with the solid violet and dashed black lines indicating the 50th, 16th, and 84th percentiles, respectively. These percentiles represent the best-fit values and associated uncertainties. In the 2D histograms, contours correspond to the $1\sigma$, $2\sigma$, and $3\sigma$ levels, with the best-fit values marked by violet points and lines. The derived values for mass, cooling age, and temperature are reported at the top of each 1D distribution panel. The negative cooling age reflects the fact that the zero point is defined at the onset of the cooling phase, implying that COM-NGC362D is still in the proto-WD phase.}
\label{fig5app}
\end{figure*}
Figure~\ref{fig5app} shows the likelihood distributions for the companion mass, effective temperature, and age of COM-NGC362D, as obtained from our multivariate Gaussian likelihood fit to the He WD models \citep[see our Sect.~\ref{section4} and Sect. 3.1 in][]{Ettorre2025a}. In the corner plot, the solid violet and dashed black lines indicate the 50th, 16th, and 84th percentiles, respectively. For each parameter, these percentiles represent the best-fit values and associated uncertainties.

\section{Pulsational instability and H-shell flashes}\label{appendix4}
\begin{figure*}[h!]
    \centering
    \includegraphics[width=0.85\hsize]{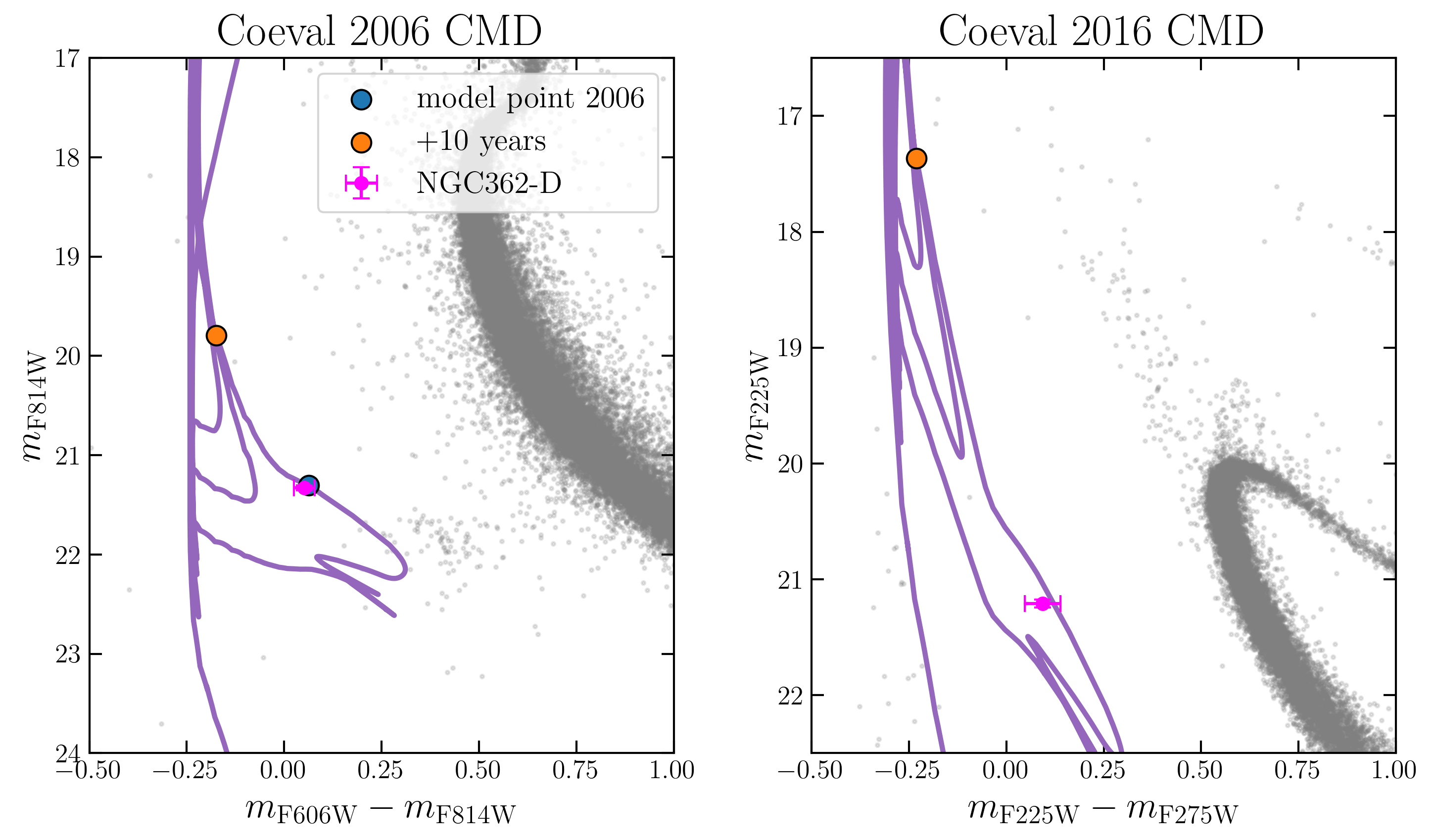}
    \caption{Colour-magnitude diagrams of NGC~362 showing the position of COM-NGC362D. \textit{Left}: ($m_{\mathrm{F606W}}-m_{\mathrm{F814W}}$, $m_{\mathrm{F814W}}$) CMD constructed from coeval observations acquired in 2006 (HST proposal GO~10775). The blue point marks the location on the cooling track that best reproduces the position of COM-NGC362D in the 2006 CMD, the orange point indicates the expected position after 10 years of evolution along the cooling track and the magenta point marks the observed position of COM-NGC362D. \textit{Right}: ($m_{\mathrm{F225W}}-m_{\mathrm{F275W}}$, $m_{\mathrm{F225W}}$) CMD combining F225W and F275W data obtained in 2016 (proposal GO-14155). In both panels, grey points represent cluster member stars and the purple line shows the $0.31\,\Msun$ ELM WD cooling track.}
    \label{fig1app}
\end{figure*}
Most WD stars experience at least one phase of pulsational instability during their evolution, during which they become multi-periodic pulsating variable stars. 
The mass of COM-NGC362D would make it belong to the ``pre-ELMV'' class of pulsating WDs, where ELMV stands for Extremely Low-Mass white dwarf Variables and refers to pulsating extremely low-mass He WDs (with masses smaller than $\sim 0.18-0.20\,\Msun$). This class of objects shows photometric variability with amplitudes of the order of 0.001–0.05 magnitude \citep{Corsico2019}. Therefore, we confidently exclude that variability associated with the instability strip plays a significant role in determining the position of COM-NGC362D in different CMDs. 

Another distinctive feature of low-mass He WDs is the occurrence of diffusion-induced hydrogen-shell flashes in the envelopes of proto-WDs with masses $\gtrsim0.20\,\Msun$. During the proto-WD phase, a substantial portion of the hydrogen retained in the envelope is consumed via stable H-shell burning. However, depending primarily on the mass, metallicity, and envelope thickness, hydrogen can also be burned through unstable CNO hydrogen-shell flashes.  Notably, WDs with thinner hydrogen layers cool considerably faster than those with thicker ones as a consequence of these flashes \citep{Istrate2016}. The mass at which hydrogen-shell flashes begin strongly depends on metallicity, and the models adopted in this letter predict that such flashes occur exclusively in WDs with masses above $\sim0.24\,\Msun$ \citep{Cadelano2020}.

Consequently, if COM-NGC362D is observed during the occurrence of one of these flashes, the use of non-coeval (random phase) observations could introduce inconsistencies between its observed position and the model predictions across different CMDs.
The left panel of Fig.~\ref{fig1app} displays the CMD built from coeval observations acquired in 2006 (HST proposal GO-10775), while the right panel shows the $m_{\mathrm{F225W}}-m_{\mathrm{F275W}}$, $m_{\mathrm{F225W}}$ CMD, combining the F225W and F275W images obtained in 2016 (proposal GO-14155). The left panel of Fig.~\ref{fig1app} also compares the position of COM-NGC362D (as a magenta point with errorbars) with the cooling track corresponding to a WD mass of $0.31\,\Msun$ (in purple), which displays three characteristic loops corresponding to three hydrogen-shell flashes. The observed position of COM-NGC362D is consistent with the source undergoing one of the flashes of the model. 

The blue point denotes the location along the $0.31\,\Msun$ cooling track that best matches the observed position of COM-NGC362D in the 2006 coeval optical CMD. The orange point indicates, in both panels, the position expected after 10 years of evolution along the same track. Under the flash hypothesis, the source observed in 2016 would therefore be expected to lie close to the orange point in the right-panel CMD.

More in general, we performed the same checks for different WD masses, in order to sample both the ascending phase of the flash, during which the magnitude changes rapidly, and the earlier descending phase of the loop, which proceeds on a much longer timescale ($\sim500$ years).  In the first case, as already shown with the $0.31\,\Msun$ model, the expected magnitude variation over 10 years is substantially larger with respect to what is observed for NGC362D. Moreover, the expected colour change increases towards redder filters, thus producing an opposite trend to what is observed. In the latter case, instead, the magnitude changes by just a few hundredths of a magnitude over 10 years, yielding a variation far too small to account for the discrepancy seen in COM-NGC362D.

\section{Photometric systematics and extinction law check}
\begin{figure*}
    \centering
    \includegraphics[width=\linewidth]{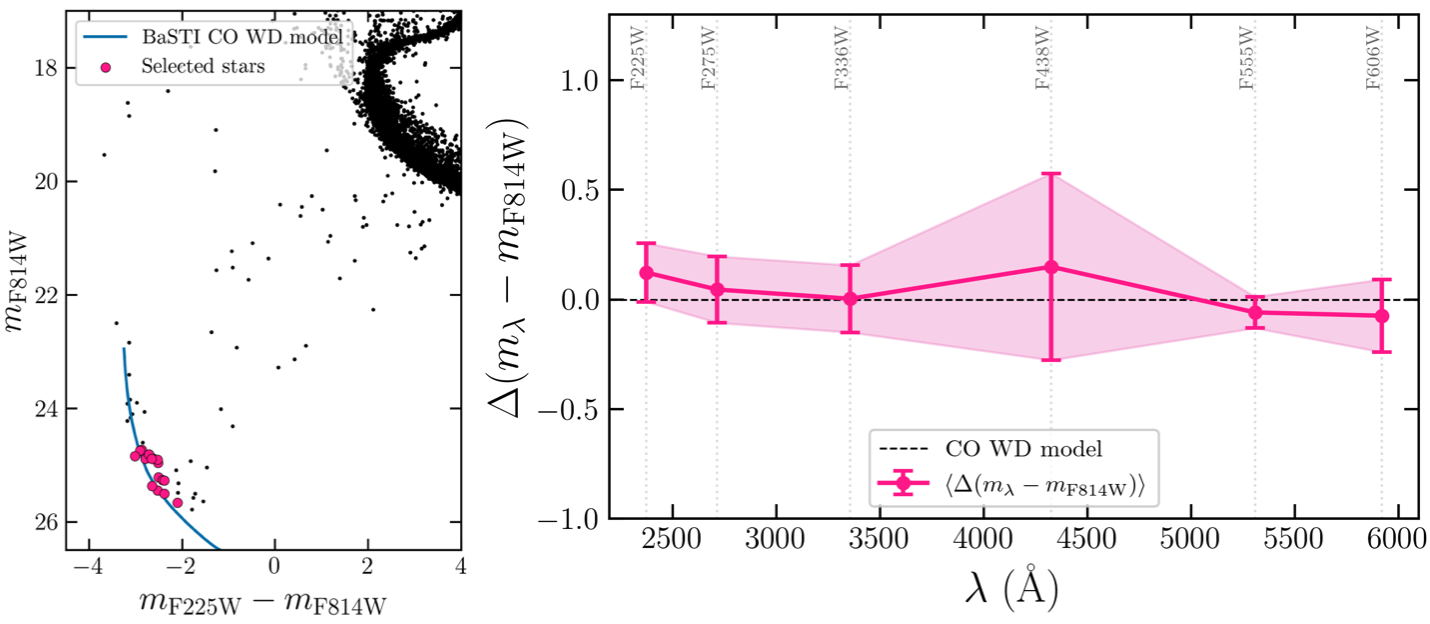}
    \caption{\textit{Left}: ($m_{\mathrm{F225W}}-m_{\mathrm{F814W}}$, $m_{\mathrm{F814W}}$) CMD centred on the CO WD region. Pink points represent the sample of 15 CO WDs in our catalogue with magnitudes available in all filters used to compute the $\Delta(m_\lambda - m_{\rm F814W})$ residuals. The blue line traces the $0.68\,\Msun$ \texttt{BaSTI} CO WD isochrone \citep{Salaris2022}. \textit{Right}: $\Delta(m_\lambda - m_{\rm F814W})$ as a function of wavelength for the selected WDs. The mean trend and its $1\sigma$ dispersion are shown by the magenta points and shaded region, respectively.}    
    \label{fig6app}
\end{figure*}
To verify that the wavelength-dependent trend shown in Fig.~\ref{fig4} is intrinsic to COM-NGC362D, we tested whether it could arise from systematic effects in the photometry or from an incorrect assumption on the adopted extinction law. We selected a sample of $\sim15$ CO WDs with magnitudes available in every filter, shown in the CMD in the left panel of Fig.~\ref{fig6app}. We computed their $\Delta_{\mathrm{m}_\lambda,\mathrm{m_{F814W}}}$ residuals with respect to a $0.68\,\Msun$ CO WD isochrone taken as a reference from the \texttt{BaSTI} database \citep{Salaris2022}. The results are shown in the right panel of Fig.~\ref{fig6app}, where we plot the mean trend (in magenta) and its $1\sigma$ dispersion. The mean residuals are consistent with zero at all wavelengths within $1\sigma$, with no significant trend, confirming that the colour excess observed for COM-NGC362D can be attributed neither to photometric systematics nor to an incorrect assumption on the extinction law.

\end{appendix}
\end{document}